\documentclass[%
 aapm,
 mph,%
 amsmath,amssymb, 
 preprint,%
]{revtex4-2} 
\usepackage[utf8]{inputenc}
\usepackage{graphicx}
\usepackage{float}
\usepackage{bm}
\usepackage{physics}
\usepackage{hyperref} 

\begin{document}
 
\title{On the Meaning of Berry Force For Unrestricted Systems Treated With Mean-Field Electronic Structure}
\author{Xuezhi Bian}
\email{xzbian@sas.upenn.edu}
\author{Tian Qiu}
\author{Junhan Chen}
\author{Joseph E. Subtonik}
\email{subotnik@sas.upenn.edu}
\date{\today}

\begin{abstract}
    We show that the Berry force as computed by an approximate, mean-field electronic structure can be meaningful if properly interpreted.  In particular, for a model Hamiltonian representing a molecular system with an even number of electrons interacting via a two-body (Hubbard) interaction and a spin-orbit coupling, we show that a meaningful nonzero Berry force emerges whenever there is spin unrestriction--even though the Hamiltonian is real-valued and formally the on-diagonal single-surface Berry force must be zero.  Moreover, if properly applied, this mean-field Berry force yields roughly the correct asymptotic motion for  scattering through an avoided crossing. That being said, within the context of a ground-state calculation, several nuances do arise as far interpreting the Berry force correctly, and as a practical matter, the Berry force diverges near the Coulson-Fisher point (which can lead to numerical instabilities).  We do not address magnetic fields here. 
\end{abstract}

\maketitle

\section{Introduction\label{sec:intro}}
Recent years have seen a dramatic rise in interest in the notion of Berry force\cite{Mead1994, Tosatti2007, Gross2010,Nagaosa2019,Nikolic2020}, i.e. the notion that classical Born-Oppenheimer (BO)  dynamics is dressed  by a  pseudo-magnetic field in many cases.
From a theoretical point of view, one of the motivations to understand Berry force is predicated on the idea that experimentally, nowadays one can produce {\em external} magnetic fields of size 30T.   In the presence of such a large magnetic field, the electronic wavefunction becomes complex-valued and a Berry force will appear and might be measurable. As such, recently Helgaker and co-workers have computed the Berry force for a system of small molecules exposed to strong magnetic fields and propagated some {\em ab initio} molecular dynamics \cite{Helgaker2021a,Helgaker2021b,Helgaker2022}. In a very different context, another drive to investigate Berry forces arises by considering how molecular motion for a small cluster can be altered when in the presence of an {\em intrinsic} magnetic fields; to that end,  Calandra and co-workers analyzed how vibrational eigenenergies are altered for small clusters of platinum \cite{Calandra2021}, and they predict measurable effects in angular momentum polarization.  

From an experimental point of view, interest in Berry force has grown recently along side the explosion of interest in the phenomenon known as chiral induced spin selectivity (CISS). As pioneered by Naaman and Waldeck \cite{Naaman2012}, and now confirmed by many others\cite{Beard2021, Naaman2019}, it is known that when a current runs through chiral molecules, one can often find a great deal of spin-selectivity -- which raises the tantalizing notion of chiral electrochemistry\cite{Naaman2015}, chiral catalysis\cite{Naaman2018}, or even chiral spintronic devices \cite{Paltiel2018}. To date, there is no satisfactory explanation for the CISS effect insofar as the relevant spin-orbit coupling (SOC) matrix elements are too small to explain the any spin filtering at room temperature. Recently, a handful of researchers \cite{Wu2020,Wu2021,Fransson2020,Fu2020,Zhang2020,Fay2021,Limmer2021,Opennheim2021} have argued that the CISS effect (which involves an {\em electronic} observable) must be tied to a breakdown of the Born-Oppenheimer approximation (which involves {\em nuclear} motion) and the emergence of a spin-dependent Berry force.

As a reminder to the reader,  the basic premise of Berry force is simple: when the Hamiltonian is either complex-valued or allows degenerate or nearly degenerate states, one must be very careful in choosing the phases of the adiabatic electronic states as a function of nuclear position; for the most part, even in a small local region of configuration space,  one cannot use parallel transport to generate a smooth gauge.   The dynamical consequences of this failure is that a pseudo-magnetic force appears which alters the BO motion.  This pseudo-magnetic force can be derived in a host of ways, e.g. by directly integrating the motion and keeping first order corrections for a mean-field force (as done by Robbins and Berry \cite{Berry1993}) or by projecting the quantum-classical Liouville equation (QCLE) (as done by our research group \cite{Subotnik2019}).   Berry forces have also been found when considering molecular junctions out of equilibrium\cite{Hedegard2010, VonOppen2012,Galperin2018}.  In any event, for motion along a discrete eigenvalue of $H_e(\vec R)$, the final  result is:
\begin{eqnarray} \label{eq:BerryForce}
\vec F_j^{\rm Berry} =  \stackrel{\leftrightarrow} { \Omega}_j  \cdot {\frac {\vec P} M}  =  2\hbar  {\rm Im}  \sum_k\left( \vec d_{jk} (\frac {\vec P} M \cdot \vec d_{kj}) \right) 
\end{eqnarray}
where $\stackrel{\leftrightarrow} { \Omega}_j = i\nabla \times \vec d_{jj}$ defines the Berry curvature of adiabatic state $\psi_j$ and  $\vec d_{jk} = \bra{\psi_j} \vec \nabla \ket{\psi_k}$ is the derivative coupling between adiabatic states $\psi_j$ and $\psi_k$. 

When inspecting the result in Eq. \ref{eq:BerryForce}, an interesting nuance arises: for a nonzero single-surface Berry force, one requires a complex-valued derivative coupling.  Now, Mead showed long ago\cite{Mead1979} that, for a system with an even number of electrons, a diabatic basis (with time-reversal) exists whereby the entire Hamiltonian is real-valued (so that one can be certain that the Berry force in Eq. \ref{eq:BerryForce} is zero). Nevertheless, for the  dynamics in Refs. \citenum{Calandra2021} one does find a nonzero Berry force for a small cluster with an even number of electrons according to a density functional theory (DFT) calculation (even with zero external magnetic field).  In other words, because of the mean-field approximation inherent to DFT, the electronic wavefunction becomes complex and a Berry force arises according to Ref. \citenum{Calandra2021}. 

At this point, one must ask: Is such a  Berry force meaningful? One could imagine arguing both sides of this argument. On the one hand, one can point out that a typical, spin-polarized unrestricted Hartree-Fock solution breaks time-reversal symmetry; accordingly, from the errors of mean-field theory, one might argue that the calculated Berry force is purely specious. On the other hand, recent work with singlet-triplet crossings (and more generally, degenerate electronic crossings) has demonstrated that Berry forces can arise even with an even number of electrons and a real-valued Hamiltonian \cite{Bian2021}.

The goal of this article is to answer these questions definitively. We will show that, if  properly interpreted, the Berry force computed from a mean-field, spin-polarized unrestricted solution {\em is} dynamically meaningful. At the same time, however, propagating dynamics along a mean-field ground state with a Berry force often highlights the incomplete nature of Born-Oppenheimer dynamics--one often misses very interesting excited state information.  Finally, as a side note,  we also show that one must  be careful when propagating mean-field dynamics with Berry forces because the Berry force will diverge at the Coulson-Fisher point.\cite{Coulson1949}

In the end, our results justify recent {\em ab initio} simulations of molecular dynamics using mean-field theory but also highlight how much more dynamical information should be recoverable if one were to propagate fully nonadiabatic {\em ab initio} dynamics in the presence of many electronic states and Berry forces. 

\section{Model\label{sec:Model}}
In order to understand how the meaning of Berry force might or might not be altered by making a mean-field electronic structure assumption, our approach will be to analyze a simple two-orbital, two-electron Hubbard model with SOC. We choose the electronic Hamiltonian as follows:
\begin{eqnarray} \label{eq:He}
    \hat H_e =\hat H_0 + \hat H_{\rm SO}  
\end{eqnarray}
\begin{eqnarray} \label{eq:H0}
    \hat H_0 = \sum_{\sigma=\uparrow,\downarrow} \left( \sum_{i=1,2}  h_{i\sigma}\hat c_{i\sigma}^\dagger \hat c_{i\sigma}  
    - t \sum_{i\neq j}  \hat c_{i\sigma}^\dagger \hat c_{j,\sigma}  \right)
    +  U \sum_{i=1,2} \hat c_{i\uparrow}^\dagger \hat c_{i\uparrow} \hat c_{i\downarrow}^\dagger \hat c_{i\downarrow} 
\end{eqnarray}
\begin{eqnarray} \label{eq:HSOC}
   \hat H_{\rm SO} = \sum_{i\neq j}  
   V_{i\sigma j \bar \sigma} \hat c_{i\sigma}^\dagger  \hat c_{j\bar \sigma}  
\end{eqnarray}
Here,  $H_0$  represents a standard Hubbard Hamiltonian where 
$\hat c_{i\sigma}$ and  $\hat c_{i\sigma}^\dagger$ are the electronic creation and annihilation operators, $h_{i\sigma}$ is the on-site energy at site $i=1,2$ with spin index $\sigma = \uparrow,\downarrow$, $t$ is the hopping term between different sites with the same electronic spin, and $U$ is the on-site Coulomb repulsion. The  term $H_{\rm SO}$  is the single electron SOC coupling where $ V_{i\sigma j \bar \sigma}$ is the coupling strength between sites $i$ and $j$ with spins $\sigma$ and $\bar \sigma$. Formally,  this operator is 
\begin{eqnarray} \label{eq:VSOCelement}
   V_{i\sigma j \bar \sigma} = \sum_{\gamma}  \xi \left( \hat L_{ij}^{\gamma} \cdot \hat S \right)_{\sigma\bar \sigma}
\end{eqnarray} 
where $\hat L^{\gamma}$ is the electronic angular momentum operator around nucleus $\gamma$ and 
$\hat S$ is the spin Pauli matrices. 
To further simplify the problem, let us imagine a model with $\hat S_z$ symmetry such that only states with the same spin in the z-direction are coupled:
\begin{eqnarray} \label{eq:VSOC}
    \hat H_{\rm SO} =  
   \sum_{i\neq j} V_{ij}  \hat c_{i\uparrow}^\dagger  \hat c_{j\uparrow} - V_{ji}  \hat c_{j \downarrow}^\dagger  \hat c_{i\downarrow}   
\end{eqnarray} 

Notice that the onsite energy $h_i = h_{i\uparrow} =h_{i\downarrow}$ and the hopping terms $t$ are real-valued variables; the SOC coupling $V = V_{ij} = -V_{ji}$ is purely imaginary (because of $\hat S_z$ symmetry). At this point, we can write the electronic Hamiltonian in  matrix form under a given basis:

\begin{equation} \label{eq:basis}
\begin{aligned}
\ket{\psi_0} &= c_{1\downarrow}^\dagger c_{1\uparrow}^\dagger \ket{0}\\
\ket{\psi_1} &=  c_{2 \downarrow}^\dagger c_{2\uparrow}^\dagger   \ket{0}\\
\ket{\psi_2} &=  c_{2 \downarrow}^\dagger c_{1\uparrow}^\dagger   \ket{0}\\ \ket{\psi_3} &= c_{1 \downarrow}^\dagger c_{2\uparrow}^\dagger   \ket{0}\\ \ket{\psi_4} &= c_{2\uparrow}^\dagger c_{1\uparrow}^\dagger   \ket{0}\\ \ket{\psi_5} &= c_{2 \downarrow}^\dagger c_{1 \downarrow}^\dagger   \ket{0} 
\end{aligned}
\end{equation} 
If we define $\tilde V \equiv -t + V$, then in this basis, the Hamiltonian reads: 
\begin{eqnarray} \label{eq:Hematrix}
\hat H_e = \begin{pmatrix}
2h_1 + U  & 0 & \tilde V & \tilde V^* & 0 & 0 \\
0 & 2h_2 + U & \tilde V  & \tilde V^* & 0 & 0 \\
\tilde V^* & \tilde V^* & h_1 + h_2 & 0 & 0 & 0 \\ 
\tilde V & \tilde V  & 0 & h_1 + h_2 & 0 & 0 \\ 
0 &0 &0 &0 & h_1 + h_2 & 0 \\ 
0 &0 &0 &0 & 0 & h_1 + h_2   \\ 
\end{pmatrix}
\end{eqnarray} 
At this point, we will allow  $h_{i}$ and $\tilde V$ to vary as functions of nuclear geometry so that we can study how the presence of SOC affects the nature of molecular dynamics. We consider a Hamiltonian that depends on  two nuclear coordinates $x$ and $y$, and we take the electronic Hamiltonian to be of the form:
\begin{eqnarray} \label{eq:h1h2}
    h_1 = - h_2 = 0.1 \tanh(x)
\end{eqnarray}
\begin{eqnarray}\label{eq:tildeV}
    \tilde V = 0.1 \exp(-\frac {x^2} 2 + iWy) 
\end{eqnarray}
\begin{eqnarray}\label{eq:U}
    U = 0.2
\end{eqnarray}
We will set $W = 5.0$ for all calculations in this paper. 
In Fig. \ref{fig:geometry}a and Fig. \ref{fig:geometry}b,
we plot the diabats (i.e. on diagonal energies) and exact adiabats (eigenvalues) for this Hamiltonian as a function of $x$.


There are three points worth mentioning about the Hamiltonian defined in Eqs. \ref{eq:Hematrix} - \ref{eq:U}.  First,  all of the diabats and adiabats are flat in the $y$ direction and all of the diabatic crossings are in the $x$ direction.  This state of affairs has been chosen on purpose to make our analysis of the dynamics easier.  In practice, as shown in Fig. \ref{fig:geometry}a and Fig.  \ref{fig:geometry}b, the adiabatic ground state  changes its diabatic composition (from diabat $\ket{\psi_0}$ when $x \to -\infty$ to two nearly degenerate diabats $\ket{\psi_2}$ and $\ket{\psi_3}$ when $x \to \infty$).  Note that, while there are two more diabats  $\ket{\psi_4}$ and $\ket{\psi_5}$ that are degenerate with diabats  $\ket{\psi_2}$ and $\ket{\psi_3}$ when $x \to \infty$, these two former states  are completely uncoupled from the latter (which again makes our interpretation easier):   we will treat the model Hamiltonian in Eq. \ref{eq:Hematrix} as if there are only four states. 
Second, the phase of the coupling $\tilde V$  between $\ket{\psi_0}$ and diabat $\ket{\psi_2}$ (and $\tilde V^*$ between $\ket{\psi_0}$ and   $\ket{\psi_3}$) is modulated in the $y$ direction by a parameter $W$.  Third,  the Hamiltonian in Eq. \ref{eq:Hematrix} can be transformed into a completely real-valued form as shown in the appendix and therefore the exact, on-diagonal Berry force (Berry curvature) formally must be zero according to Eq.\ref{eq:BerryForce}. 

The focus of this article is understanding the consequences of the third point above. Recently,  our group has shown\cite{Miao2019,Bian2021}  that a ``Berry-like'' force can be important in model systems like Eq. \ref{eq:Hematrix}.   For instance,  the complex-valued phase modulation $e^{iWy}$ in Eq. \ref{eq:tildeV} causes a momentum shift of $W$ in the $y$ direction for nuclear trajectories that transmit through the crossing adiabatically \cite{Miao2019, Bian2021}.  Moreover, the nuclear dynamics will be changed dramatically if the magnitude of $W$ is comparable to nuclear momentum $\vec P$. But how can ``Berry-like'' forces arise with a formally zero  on-diagonal single-surface Berry force?
One must conclude that these effects can be captured only with a nonadiabatic (multi-state) formalism.  To that end, recently we have sought to explain such features using a quasi-diabatic formulation of fewest-switches surface-hopping (FSSH) dynamics as appropriate for a singlet-triplet crossing \cite{Bian2022}, where we can force the Berry force to be nonzero. From a more general point of view, we have also proposed a phase-space formulation of the FSSH that by design  takes into account the multi-dimensional nature of degenerate Berry curvature tensor\cite{Wu2022}. 

Now, in considering all of the formal theory above, one important twist is missing: all of the formal, dynamical theory of Berry forces is predicated on the assumption that one has knowledge of the exact Hamiltonian eigenstates,  eigenforces, and the derivative couplings.  And in such a formal case, as emphasized above repeatedly, the on-diagonal Berry force corresponding to any exact eigenstate is zero.  That being said, however, this formal result need not hold for such a system with a reduced electronic structure description. As discussed in the introduction, for a realistic electronic Hamiltonian with electron-electron interactions (as in Eq. \ref{eq:Hematrix}), if one performs a spin unrestricted Hartree-Fock or DFT calculation, the ground-state electronic wavefunction will be one of a pair of broken symmetry solutions and therefore likely be complex-valued with a nonzero Berry curvature. Will such a Berry curvature be meaningful? 
In practice, single Slater determinant based mean-field methods such as DFT and HF are dominant in the field of electronic structure calculations of Born-Oppenheimer dynamics, and so answering this question is quite important going forward. 

\subsection{Mean Field Theory}
We begin with the details of a mean-field generalized Hartree-Fock (GHF) calculation.
The system is represented in a single-electron spin basis:
\begin{eqnarray}\ket{\chi_{i\sigma}} = \hat c_{i\sigma}^\dagger \ket{0}
\end{eqnarray} 
Then, each spinor orbital is written as a linear combination
of these basis functions:
\begin{eqnarray}
    \ket{\phi_k} = \sum_{\sigma=\uparrow,\downarrow}  \sum_{i=1,2} 
    C_{i\sigma,k}  \ket{\chi_{i\sigma}}
\end{eqnarray}
Here $C_{i\sigma,k}$ are the molecular orbital (MO) coefficients. The Fock operator can be expressed as:
\begin{eqnarray} \label{eq:Fock}
    \hat F = \hat h + \hat G  
\end{eqnarray}
where in the matrix form, the single-electron term $\hat h$ and two-electron term $\hat G$ are:
\begin{eqnarray} \label{eq:1eletron}
    \hat h = \begin{pmatrix}
    h_1  & \tilde V & 0 & 0 \\ 
    \tilde V^*  & h_2 & 0 & 0 \\ 
    0 & 0 & h_1  & \tilde V^* \\ 
    0 & 0 & \tilde V  & h_2  
    \end{pmatrix}
\end{eqnarray}
\begin{eqnarray}\label{eq:2seletron}
    \hat G = \begin{pmatrix}
     U\langle n_{1,\downarrow}\rangle & 0 & 0 & 0 \\ 
    0  & U\langle  n_{2,\downarrow}\rangle& 0 & 0 \\ 
    0 & 0 &  U\langle n_{1,\uparrow}\rangle & 0 \\ 
    0 & 0 &  0 &  U\langle n_{2,\uparrow}\rangle
    \end{pmatrix}
\end{eqnarray}
The electron density  $\langle n_{i\sigma} \rangle$ is calculated by $\langle n_{i\sigma} \rangle = \rho_{i\sigma i\sigma}$, where the density matrix $\hat \rho$ is defined as:
\begin{eqnarray}
    \rho_{i\sigma j\bar\sigma} = \sum_{k=1}^{2}  C_{i\sigma,k} C^*_{j\bar \sigma,k}
\end{eqnarray}
Next, the system is solved self-consistently by a complex Roothann equation:
\begin{eqnarray}
    FC = \varepsilon C
\end{eqnarray}
with the total energy:
\begin{equation}
    E = \frac 1 2 \left[ \Tr(\hat h \hat \rho^\dagger) + \Tr(\hat F \hat \rho^\dagger) \right]
\end{equation}
The ground state wavefunction is approximated to be a single Slater determinant of the form:
\begin{eqnarray} \label{eq:phi0}
    \ket{\Phi_0} = \frac 1 {\sqrt 2} 
    \left(\ket{\phi_1(\vec r_1)\phi_2(\vec r_2)} - \ket{\phi_1(\vec r_2)\phi_2(\vec r_1)}\right)
\end{eqnarray}
The ground state Berry curvature is
\begin{eqnarray} \label{eq:Berrycurvature}
    \stackrel{\leftrightarrow} { \Omega}_0  = i\vec \nabla \times \bra{\Phi_0} \vec \nabla \ket{\Phi_0}  = i( \vec \nabla \times   \bra{\phi_1} \vec \nabla  \ket{\phi_1} + \vec \nabla \times \bra{\phi_2} \vec \nabla  \ket{\phi_2}) \
\end{eqnarray}
or in index form,
\begin{eqnarray} \label{eq:Berrycurvatureindex}
    \Omega_0^{xy} =  \sum_{j=1}^2 i(\nabla_x  \bra{\phi_j}  \nabla_y  \ket{\phi_j} -\nabla_y \bra{\phi_j}  \nabla_x  \ket{\phi_j})
\end{eqnarray} 
Note that the GHF ground state $\ket{\Phi_0} $ is always two-fold degenerate  Kramers pair that can be found by applying the time reversal operator, $ \mathcal{T}$:
\begin{eqnarray}
    \ket{\Phi_0'} = \mathcal{T} \ket{\Phi_0} 
\end{eqnarray}
The Berry curvature of $\ket{\Phi_0}$ and $\ket{\Phi_0'}$ are related by time reversal symmetry as well,
\begin{eqnarray}   \stackrel{\leftrightarrow} {\Omega'}_0 = \mathcal{T} \stackrel{\leftrightarrow} {\Omega}_0  \mathcal{T}^{-1} =  -\stackrel{\leftrightarrow} {\Omega}_0 
\end{eqnarray}

\subsection{Potential Energy Surfaces}
\begin{figure} [H]
    \includegraphics[width=1\columnwidth]{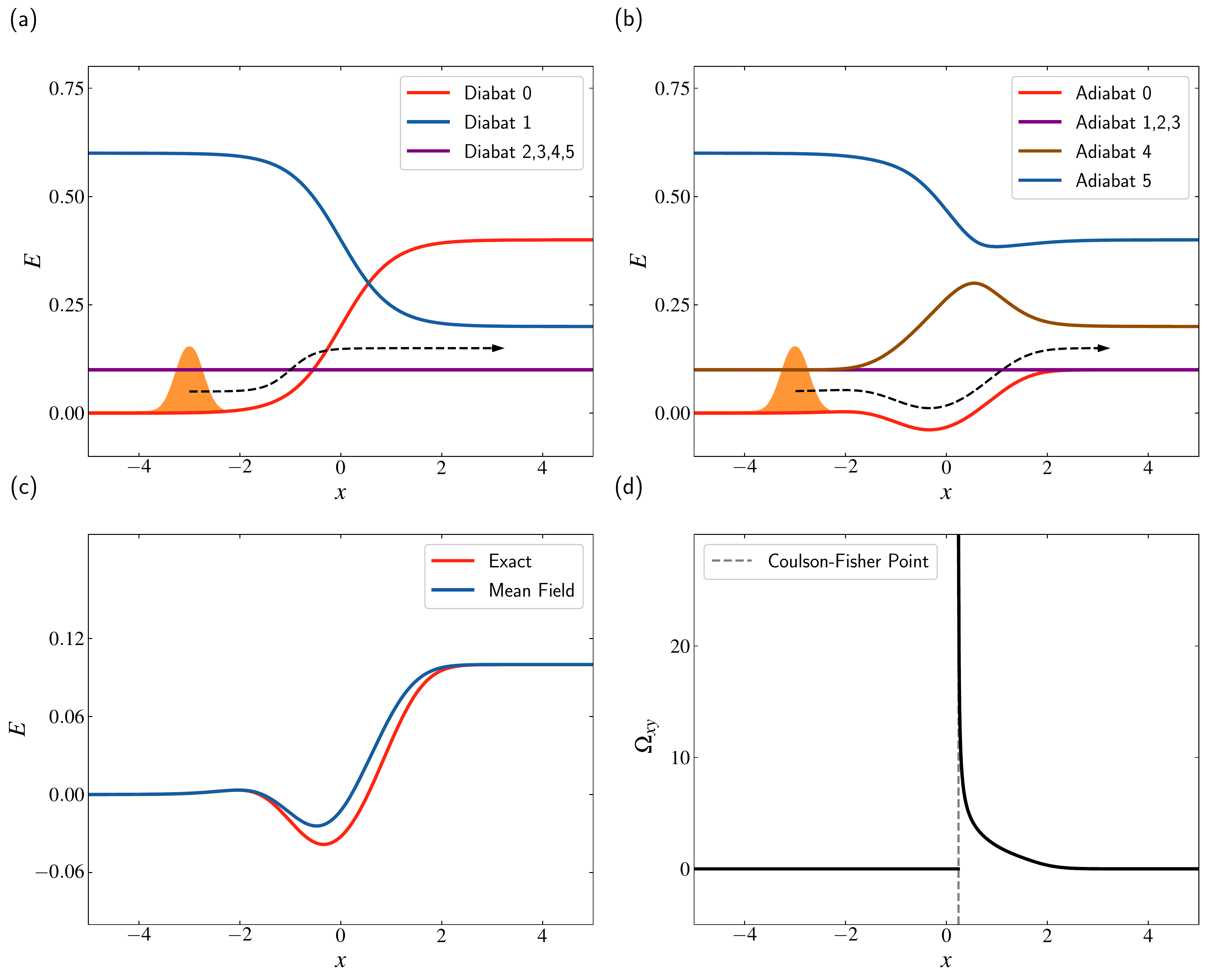}
    \caption{\label{fig:geometry} (a),(b) The exact diabatic and adiabatic
    PESs of Hamiltonian described by Eq. \ref{eq:Hematrix} along the $x$ direction. (c) Zoomed-in plot of the ground state PESs as calculated from exact diagonalization and a mean field GHF ansatz. (d) The ground state Berry curvature (for $\ket{\Phi_0}$ [see Eq. \ref{eq:phi0}]) as calculated from the mean field method. The Berry curvature diverges at the Coulson-Fisher point $x_{\rm C-F}$ near $x = 0$.}   
\end{figure}

A zoomed-in plot of the exact versus mean-field ground state potential energy surface (PES) is given in Fig. \ref{fig:geometry}c. Note that the mean-field solution is not smooth; there is a Coulson-Fisher point $x_{\rm C-F}$ near $x = 0$. A paired GHF solution is preferred when $x < x_{\rm C-F}$ and  an unpaired GHF solution is preferred when $x > x_{\rm C-F}$.  

The ground state Berry curvature has been calculated according to Eq. \ref{eq:Berrycurvature} by finite difference. In Fig. \ref{fig:geometry}d, we see that the Berry curvature is zero on the paired side ($x < x_{\rm C-F}$)  since the wavefunction is completely real-valued. However, the Berry curvature is non-zero on the unpaired side ($x > x_{\rm C-F}$); the Berry curvature diverges at the Coulson-Fisher point $x_{\rm C-F}$.  

\section{ Results \label{sec:result}}
To assess the validity of the mean-field Berry force, we have run two different sets of scattering calculations. In the first approach, we simulated exact quantum dynamics for the Hamiltonian in Eq. \ref{eq:Hematrix} using the fast Fourier transform split operator method\cite{Kosloff1983}. We initialized a Gaussian wave packet on the diabat $\ket{\psi_0} = \hat c^{\dagger}_{1\downarrow}c^{\dagger}_{1\uparrow}\ket{0}$ as:
\begin{eqnarray} \label{eq:initwf}
 \label{eq:wp} \ket{\Psi({\vec R})} =  \exp \left(-\frac {({\vec R} - {\vec R_0})^2} {\sigma^2} +  \frac {i{\vec P}_0 \cdot {\vec R}} {\hbar}  \right) \ket{\psi_0} 
\end{eqnarray} 
with initial position $\vec R_0 = (-3,0)$ and  initial momentum $\vec P_0 = (P_{\rm init}^x, 0)$. We set the wave packet width in the position space to be $\sigma_x = \sigma_y = 1$ and the nuclear mass to be $M = 1000$, and then propagated four electronic states over two nuclear dimensions.

In the second approach, we simulated Born-Oppenheimer dynamics on the mean-field ground state surface. We calculated and fit the ground state PES $E_0$ and Berry curvature $\Omega_0$ along the $x$ direction with $\Delta x = 2 \times 10^{-4}$; a linear interpolation was used during the propagation to reduce the numerical instability and computational cost.
Then we sampled $10^3$ classical trajectories according to the Wigner distribution of Eq. \ref{eq:initwf}. The trajectories were  propagated using a standard velocity Verlet method\cite{Verlet1967} with equation of motion:
\begin{eqnarray}
 \dot {\vec R} = \frac {\vec P} M 
\end{eqnarray}
\begin{eqnarray}
 \dot {\vec P} = -\vec \nabla E_0 + \vec F_0^{\rm Berry}  
\end{eqnarray}
where we set $\vec F_0^{\rm Berry} = (\stackrel{\leftrightarrow} { \Omega}_0  \cdot   {\vec P} )/ M$ for BO dynamics with Berry force  and $\vec F_0^{\rm Berry} = 0$ for BO dynamics without Berry force.

\begin{figure} [H]
    \includegraphics[width=1\columnwidth]{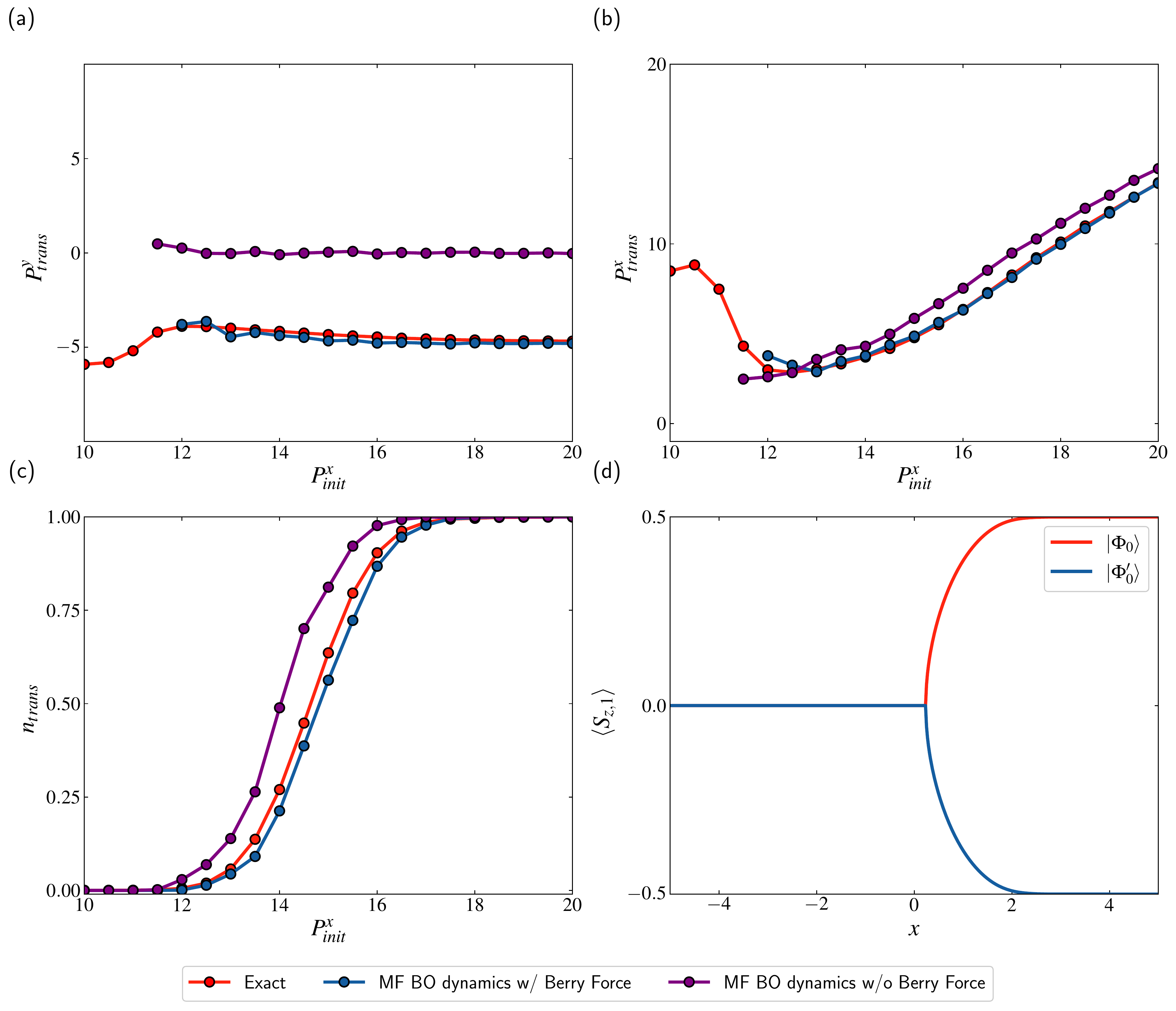}
    \caption{\label{fig:comparedynamics}
    (a) The final transmitted momentum $P_{\rm trans}^y$ in the $y$ direction as calculated by (1) exact wave packet dynamics on $\ket{\psi_{2}}$; (2)  BO mean-field ground state trajectories without a Berry force;  (3) BO dynamics with Berry force as calculated from GHF solution $\ket{\Psi_0}$.  
    (b) The final momentum $P_{\rm trans}^x$ in the $x$ direction. Note that one finds ``missing points'' for BO dynamics in (a) and (b) because no classical trajectory can transmit at low momentum.
    (c) The comparison of total transmission populations $n_{\rm trans}$.  Note that, because we set $P_{init}^y = 0$, we recover the exact same transmission results if we employ the Berry force from either GHF solution $\ket{\Psi_0}$ or GHF solution $\ket{\Psi_0'}$.
    (d) The local spin at site 1 $\langle S_{z,1} \rangle$ for two GHF solutions that distinguishes the broken symmetry solutions $\ket{\Phi_0}$ from $\ket{\Phi_0'}$. Altogether, BO dynamics with Berry force recapitulate exact quantum results quite well.  
    }
\end{figure}

Our numerical results are presented as follows in Fig. \ref{fig:comparedynamics}.  

We begin by considering the exact dynamics. For conditions that begin on diabat $\ket{\psi_0}$ (see Eq. \ref{eq:basis}) with low incoming energy starting at $x=- \infty$,  the scattering wavepacket can branch onto only two surfaces ($\ket{\psi_2}$ and $\ket{\psi_3}$);  at higher energy, four states are possible.  
After the scattering event, each wavepacket picks up a momentum with a magnitude $\pm W$ in the $y$ direction ( $+W$ for the wave packet that ends up on diabat $\ket{\psi_2}$ and $-W$ for the wave packet that ends up on the other diabat $\ket{\psi_3}$); recall that no density can emerge on diabats $\ket{\psi_4}$ and $\ket{\psi_5}$ as these diabat states are uncoupled.
In Fig. \ref{fig:comparedynamics}a and Fig. \ref{fig:comparedynamics}b, we show the final momenta $P_{\rm trans}^y$ and $P_{\rm trans}^x$ for the transmitting wave packet on diabat $\ket{\psi_2}$ according to exact dynamics. 

Next, let us consider BO dynamics along the mean-field ground state surface.   The interesting feature of the BO dynamics is that one passes through the Coulson-Fisher point around $x_{\rm C-F} \approx 0$;  we find that the (spurious) infinite magnetic field  can be integrated out safely and converged.
After passing through the Coulson-Fisher point, there are two different unpaired mean-field GHF solutions to the electronic Hamiltonian in Eq. \ref{eq:Fock}-\ref{eq:2seletron}. These two different solutions can be characterized by their respective $\langle S_{z} \rangle$ values.\cite{footnote1}
Here, we define the local spin $\langle S_i \rangle$ at site $i$ as:
\begin{eqnarray} \label{eq:Sz}
\langle S_i \rangle =   \frac 1 2
\begin{pmatrix}  C^*_{i,\uparrow} & C^*_{i,\downarrow}   \end{pmatrix}
 \hat \sigma_z
\begin{pmatrix}  C_{i,\uparrow} \\ C_{i,\downarrow}   \end{pmatrix}
\end{eqnarray}
In Fig. \ref{fig:comparedynamics}d, we plot the $\langle S_{z,1}\rangle$ values of the two GHF solutions $\ket{\Phi_0}$ and $\ket{\Phi'_0}$. In this ``unpaired region'' when $x > x_{\rm C-F}$, the values of Eq. \ref{eq:Sz} are equal and opposite for the two different mean-field solutions. In Fig. \ref{fig:comparedynamics}a, we show that, if we run dynamics using the ground state with Berry curvature as calculated for the mean-field solution characterized by $\ket{\Phi_0}$ (in Fig. \ref{fig:comparedynamics}d), the resulting observables match exact quantum dynamics for transmission to diabat $\ket{\psi_2}$. In particular, trajectories can pick up a quantitatively correct shift in momentum with consideration of Berry force.  Note that, if we were to run dynamics with the mean-field solution characterized by $\ket{\Phi_0'}$, we would match the asymptotic momentum on  diabat $\ket{\psi_3}$.
Thus, a lot of the correct physics present within a mean-field framework; one must be simply be careful in how one interprets the mean-field result. This general conclusion is in agreement with recent studies of electron-phonon interactions in a junction and negative differential resistance.\cite{Dou2016,Galperin2004}

Finally, we note that the existence of nonzero Berry force changes the energy barrier for the scattering event.  More precisely, we find that the effective energy barrier rises because Berry force leads to  energy being shared between the $x$ and $y$ degrees of freedom, and energy conservation then demands that the barrier be raised in the $x-$direction.  One can quantify this change by considering the transmission/reflection ratio with and without Berry force, as shown in Fig.\ref{fig:comparedynamics}c. Clearly,  mean-field BO dynamics are improved by including Berry force calculations, where we see less transmission over the range $14 < P_{init}^x < 16$.

\begin{figure} [H]
    \includegraphics[width=1\columnwidth]{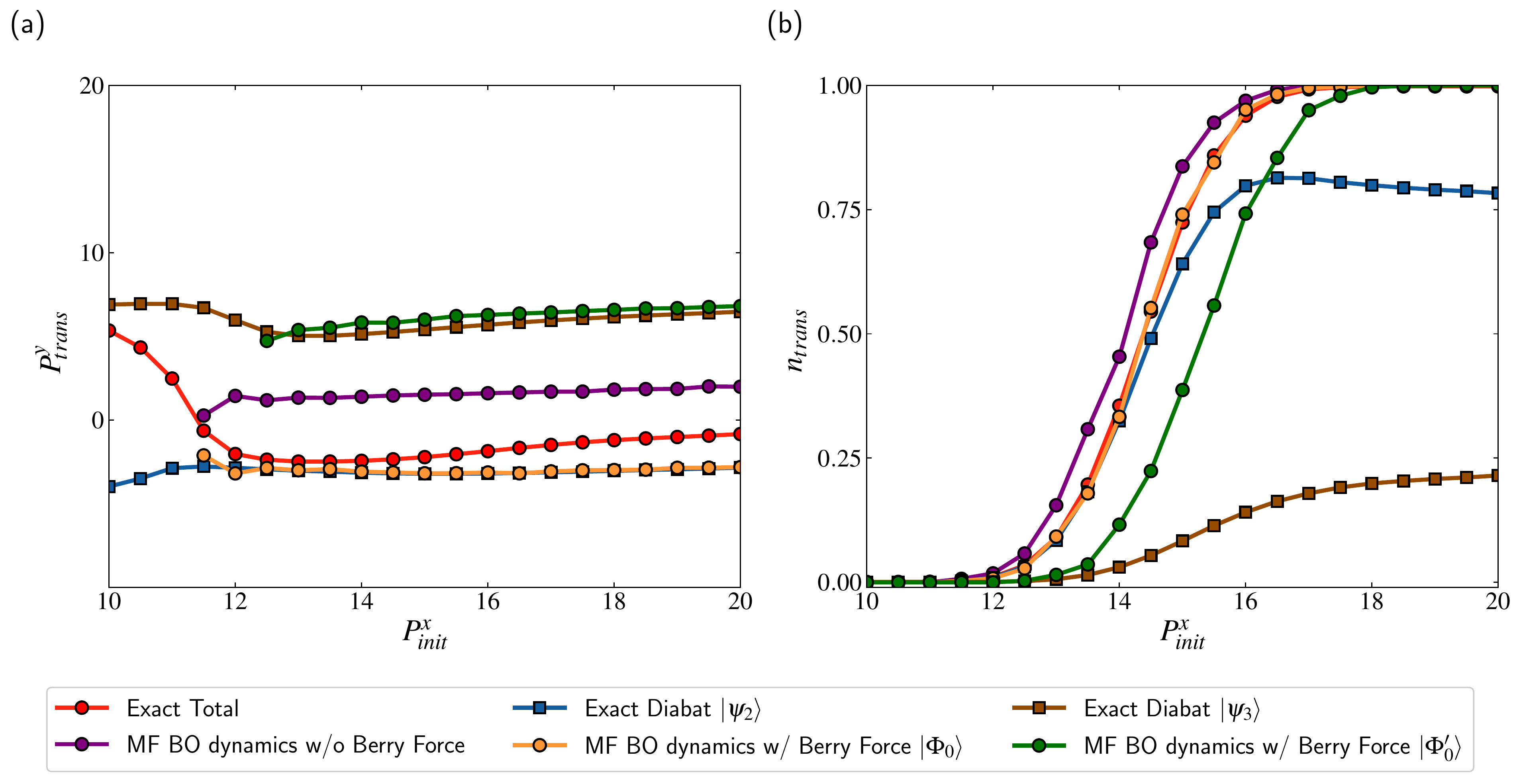}
    \caption{\label{fig:comparedynamicsangle}
    The result of dynamics similar to Fig. \ref{fig:comparedynamics}$a,c$, but now the system starts with initial momentum $P_{init}^y =  0.1P_{init}^x$. 
    For BO dynamics, we show results for both GHF solutions (which have equal and opposite Berry forces).  We compare against exact results, where momenta and populations on diabats $\ket{\psi_2}$ and $\ket{\psi_3}$ are shown separately (along with the overall [average] results). In $(a)$, note that one can capture the asymptotic $y-$momentum of each channel with BO results, but the total $y-$ momentum cannot be retrieved.  In $(b)$, note that the transmission probabilities are very different for dynamics along $\ket{\Phi_0}$ vs $\ket{\Phi_0'}$. In particular, if one performs a calculation with $\ket{\Phi_0}$, one can match the total transmission probability; but if one were to rely on a calculation with $\ket{\Phi_0'}$, one would recover an erroneous transmission probability.
    }
\end{figure}

\section{Discussion and Conclusions\label{sec:discussion}}
The conclusion of this article is that, whenever a system is described by a spin-unpaired GHF state that breaks time-reversal symmetry, the Berry force as computed with running {\em ab initio} mean-field theory is meaningful.
One must simply be aware, however, that this Berry force is one of a pair of Berry forces and there is an equal and opposite  Berry force in the opposite direction attached to the other degenerate state found by time reversibility. Although we have arrived at this conclusion by considering a problem with degenerate electronic states (at infinity), the same conclusion should hold for any unpaired solution (no matter how big is the fundamental energy gap). Thus, there is clearly some merit to running Born-Oppenheimer dynamics calculations on {\em ab initio} surfaces with {\em ab initio} Berry forces -- even when formally the exact single-state Berry force should be zero.

That being said, however, before investing in massive {\em ab initio} calculations,  several key questions should really be addressed in principle. First,  here we have addressed a scattering calculation for which two equal and opposite wave packets are spawned when the molecule approaches a crossing and goes through a Coulson-Fisher point. What would be the implications if we were to treat a bound-state problem where wave packets cannot escape? For instance,  would the effect of Berry force on the simulations  of vibrational eigenstates  be meaningful when the electronic structure is taken as mean-field theory. We cannot yet be certain but if so, one has all the more reason to run {\em ab initio} nonadiabatic dynamics with Berry forces, as in Ref. \citenum{Wu2022}.

Second, the data in Fig. \ref{fig:comparedynamics} makes the simplifying approximation that the momentum of the initial wave packet is entirely in the $x$ direction (i.e. $P_{\rm init}^y = 0$).  For this Hamiltonian, although not shown above, one would recover the same transmission function (in Fig. \ref{fig:comparedynamics}) for either of the two unrestricted mean-field solutions.  Unfortunately, however, if one initializes the dynamics with $P_{\rm init}^y \ne 0$, the situation becomes more complicated. For example, consider Fig. \ref{fig:comparedynamicsangle}, where we plot the dynamics for 
$P_{init}^y =  0.1P_{init}^x$.  In Fig. \ref{fig:comparedynamicsangle}a, we plot the $y-$momentum ($P^y_{\rm trans}$) as calculated from BO dynamics with two GHF Berry forces.  As one might hope, the latter quantities do agree with  corresponding exact dynamics on two diabats: $\ket{\psi_2}$ matches $\ket{\Phi_0}$ and  $\ket{\psi_3}$ matches $\ket{\Phi_0'}$.  The total transmitted $y$-momentum (in red) is a weighted average of the latter two quantities.  Note that such a weighted average cannot be simply computed with only  a pair of BO calculations; if one were to average the results from BO simulations with 50\% of trajectories experiencing the Berry force from $\ket{\Phi_0}$ and 50\% experiencing the Berry force from  $\ket{\Phi_0'}$, one would find approximately the purple curve (which is the result without any Berry force at all). One must know the different probabilities of populating each channel.  Without such probabilities, one cannot match the exact transmitted $y$-momentum (in red).

Next, consider Fig. \ref{fig:comparedynamicsangle}b where we plot the probability of transmission $n_{\rm trans}$. Here, we find the transmitted wavepacket population distribution between diabats $\ket{\psi_2}$ and $\ket{\psi_3}$ changes with the initial momentum $\vec P_{init}$. Surprisingly, we find the the total probability of transmission (in red) matches with the BO dynamics with Berry force from GHF solution $\ket{\Phi_0}$ (in orange). This quantity is different from the transmission probability as calculated with the Berry force from the GHF solution $\ket{\Phi_0'}$. This difference can be rationalized by realizing that, according to exact dynamics, at low momentum the transmission is mostly on diabat $\ket{\psi_2}$. 
At higher momentum ($P_{init}^x>16$), both BO trajectories (with either $\ket{\Phi_0}$ or $\ket{\Phi_0'}$ ) transmit. Therefore, the orange curve (for $\ket{\Phi_0}$) effectively matches the cumulative exact curve in red -- even though there are considerable amount of population on $\ket{\psi_3}$. Note that the exact opposite scenario would unfold (where the green curve would match the red curve) were we to initialize all dynamics with $P_{init}^y = -0.1P_{init}^x$.  As above, it would seems that one can extract a reasonable amount of information from a BO calculation when properly interpreted, especially if supplemented with experimental or high level branching ratios. 
Without any supplementation information, there are limitations to BO dynamics and in many cases, a  nonadiabatic, multi-state simulation is preferred.
 
\begin{figure} [H]
    \includegraphics[width=1\columnwidth]{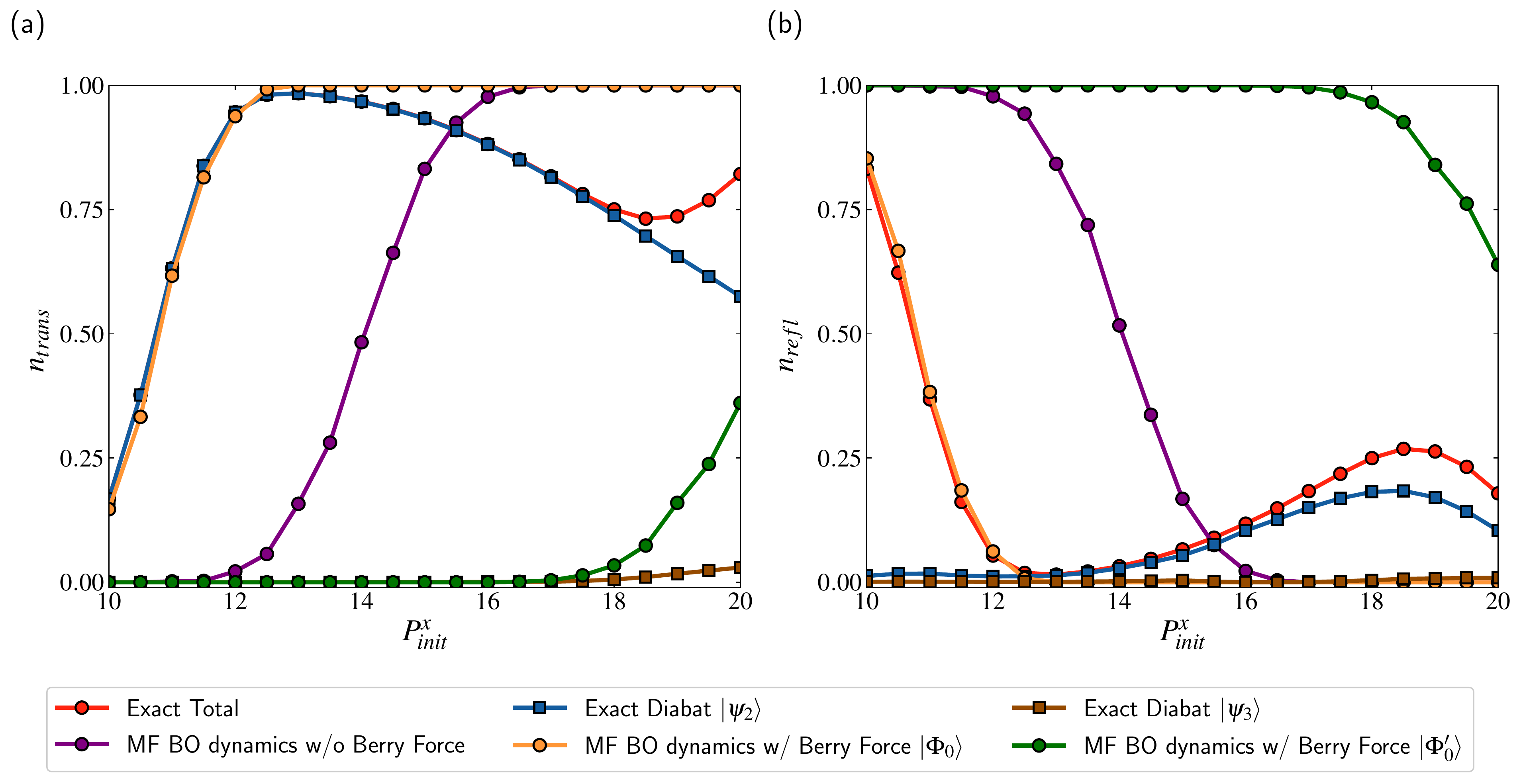}
    \caption{\label{fig:comparedynamicsangle2}
    The system is initialized the same as in Fig.\ref{fig:comparedynamicsangle} except with the initial momentum $P_{init}^y =  P_{init}^x$. Both transmission and reflection are plotted. Note that no BO calculation can match the exact transmission data for $P_{init}^x > 14$; in this range, the dynamics are  in the nonadiabatic regime. 
    }
\end{figure}
 
This last point can be made more explicit by considering a slightly different dynamical situation. In Fig. \ref{fig:comparedynamicsangle2}, we plot transmission and reflection probabilities where we 
initialize the system with  $P_{init}^y =  P_{init}^x$ (as opposed to  $P_{init}^y =  0.1P_{init}^x$ in Fig. \ref{fig:comparedynamicsangle}).  For small values of $P_{init}^x$,  the dynamics prefer transmission on  diabat state $\ket{\psi_2}$ and the total transmission can be calculated with BO dynamics according to the GHF solution $\ket{\Phi_0}$.  However, this approximation breaks for $P^x > 14$, where according to exact dynamics there is an increasing amount of reflection; no such trend is observed with BO dynamics.   This phenomenon can be understood by realizing that  the system becomes more and more nonadiabatic as $P_y$ increases;  in such a limit, the value of a BO calculation becomes very limited and a nonadiabatic simulation in the spirit of Ref. \citenum{Wu2022} is far more useful.

Interestingly, the framework above has some relevance as far as considering the dynamics of a molecule on a metal surface. Whenever a (slow) molecule interacts with a metal (with fast electrons), the molecular motion experiences both drag and a random force; the drag is often referred to as the electronic friction tensor (for which there is a long literature going back to Suhl\cite{Suhl1975} and then Head-Gordon and Tully\cite{Tully1995}).  Recently,  we have argued that, whenever the Hamiltonian is complex-valued representing a molecule on a metal surface with spin orbit coupling, the electronic friction tensor can become very asymmetric as a  large Berry force emerges\cite{Teh2021}. We have further hypothesized that this force may be responsible for CISS physics\cite{Teh2021:spin}.  Given that mean-field theory can offer a Berry force with some validity, it would now appear prudent to calculate the asymmetric component of the electronic friction tensor within DFT, and ascertain whether any new emergent spin physics emerges.

Finally, it is worth noting that most of the interesting physics in this model problem centers around the Coulson-Fisher point in Fig. \ref{fig:geometry}d that  arises when one has a big $U$ term and the preferred ground state switches from closed to open shell. At the Coulson-Fisher point, the Berry force diverges and the first derivative of the ground state energy becomes not smooth (and a related divergence is sometimes found in the context of the symmetric friction tensor describing molecular dynamics on a metal surface\cite{persson:2005:friction_noncondon,persson:2007:friction_noncondon}.)  For the most part, one is usually very hesitant to run dynamics with GHF wave functions because running dynamics can be unstable and/or chaotic when PESs are not smooth. And yet, here we find the divergence of $F^{Berry}$ does not ruin the trajectory (provided we use a small enough time step [which might or might not be tedious]) and may even help recover the correct asymptotic physics given the limitations of mean-field theory. Future work will be necessary as far as understanding if other unanticipated practical problems  arise when running dynamics around Coulson-Fisher points for large systems. As a practical matter, one can also anticipate that future work will also  investigate new techniques in electronic structure that can capture static correlation for systems with spin-orbit coupling.
Indeed, the future of {\em ab initio} nonadiabatic dynamics with Berry forces would appear to be a very exciting new research direction. 


\appendix
\section{\label{sec:appendix}}
Here we show that the exact Hamiltonian in Eq. \ref{eq:Hematrix4} can be made completely real-valued by a unitary transformation. The Hamiltonian is:
\begin{eqnarray} \label{eq:Hematrix4}
\hat H_e = \begin{pmatrix}
2h_1 + U  & 0 & \tilde V & \tilde V^*  \\
0 & 2h_2 + U & \tilde V  & \tilde V^*  \\
\tilde V^* & \tilde V^* & h_1 + h_2 & 0  \\ 
\tilde V & \tilde V  &  0 & h_1 + h_2 \\ 
\end{pmatrix} 
\end{eqnarray} 
This Hamiltonian can be transformed as follows:
\begin{eqnarray} 
\hat H_e' = R^\dagger \hat H_e R = \begin{pmatrix}
2h_1 + U  & 0 & \sqrt{2} \Re[\tilde V] & \sqrt{2} \Im[\tilde V]  \\
0 & 2h_2 + U & \sqrt{2} \Re[\tilde V] & \sqrt{2} \Im[\tilde V] \\
\sqrt{2} \Im[\tilde V]  & \sqrt{2} \Im[\tilde V]  & h_1 + h_2 & 0  \\ 
\sqrt{2} \Re[\tilde V] & \sqrt{2} \Re[\tilde V] &  0 & h_1 + h_2 \\ 
\end{pmatrix} 
\end{eqnarray} 
where
\begin{eqnarray}
R = \begin{pmatrix}
1 & 0 & 0 & 0  \\
0 & 1 & 0 & 0 \\
0 & 0 & \frac 1 {\sqrt{2}} & \frac 1 {\sqrt{2}} \\ 
0 & 0 & \frac {-i} {\sqrt{2}} & \frac i {\sqrt{2}} \\ 
\end{pmatrix} 
\end{eqnarray}

\bibliography{main.bib}

\begin{thebibliography}{45}
\bibitem [{foo()}]{footnote1}%
  \BibitemOpen
  \href@noop {} {}\bibinfo {howpublished} {Note that, in the basis
  Eq.\ref{eq:basis}, the total $\hat S_z$ operator takes the form $\hat S_z$ =
  0. Thus, $S_z$ is a good quantum number and identically zero for all states
  of interest.}\BibitemShut {Stop}%
\end{thebibliography}
\end{document}